\documentclass[ times,3p,sort&compress ]{elsarticleTCN}

\usepackage{amsfonts,amssymb,latexsym,xspace,epsfig,graphicx,color}
\usepackage{amsmath,enumerate,stmaryrd,xy,stackrel}

\usepackage[latin9]{inputenc}
\usepackage[T1]{fontenc}
\usepackage[english]{babel}
\usepackage{subfigure}
\usepackage{cancel}
\usepackage{ulem}
\usepackage{amssymb}
\usepackage{epsfig}
\usepackage{appendix}
\usepackage{bm}
\usepackage{dcolumn}
\usepackage{multirow}
\usepackage{float}
\usepackage{listings}
\usepackage{fancyhdr}
\usepackage[mathscr]{eucal}
\usepackage{ulem}
\usepackage{color}
\definecolor{SLvermilion}{cmyk}{0,.8,1,0}
\definecolor{SLorange}{cmyk}{0,0.5,1.0,0}
\definecolor{SLblue}{cmyk}{1.0,0.5,0,0}
\def\colorByNumerA#1{%
   \ifnum #1=4\relax\color{SLblue}
   \else\ifnum #1=5\relax\color{SLorange}
   \else\ifnum #1=2\relax\color{SLvermilion}\fi\fi\fi}
\include{srctex}

\begin{document}

\begin{frontmatter} 
\title{Tipping points in opinion dynamics: a universal formula in five dimensions}

\author[sg]{Serge Galam}
\ead{serge.galam@sciencespo.fr} 
\author[tc]{Taksu Cheon}
\ead{taksu.cheon@kochi-tech.ac.jp} 

\address[sg]{
CEVIPOF - Centre for Political Research, Sciences Po and CNRS,
98 rue de l'Universit\'e, 75007 Paris, France} 
\address[tc]{
Laboratory of Physics, Kochi University of Technology, 
Tosa Yamada, 
Kochi 782-8502, Japan} 

\date{May 26, 2020}


\begin{abstract}

A universal formula is shown to predict the dynamics of public opinion including eventual sudden and unexpected outbreaks of minority opinions within a generic parameter space of five dimensions. The formula is obtained combining and extending several components of Galam model of opinion dynamics, otherwise treated separately, into one single update equation, which then deploys in a social space of five dimensions. Four dimensions account for a rich diversity of individual traits within a heterogeneous population, including differentiated stubbornness, contrarianism, and embedded prejudices. The fifth dimension is  the size for the discussing update groups. Having one single formula allows exploring the complete geometry of the underlying landscape of opinion dynamics. Attractors and tipping points, which shape the topology of the different possible dynamics flows, are unveiled. Driven by repeated discussions among small groups of people during a social or political public campaign, the phenomenon of minority spreading and parallel majority collapse  are thus revealed ahead of their occurrence. Accordingly, within the opinion landscape, unexpected and sudden events like Brexit and Trump victories become visible within a forecast time horizon making them predictable. Despite the accidental nature of the landscape, evaluating the parameter values for a specific case allows to single out which basin of attraction is going to drive the associate dynamics and thus a prediction of the outcome becomes feasible. The model may apply to a large spectrum of social situations including voting outcomes, market shares and societal trends, allowing to envision novel winning strategies in competing environments.

\end{abstract}
\end{frontmatter}

\section{Introduction}
Majoritarian social decisions have been traditionally  justified by Condorcet's jury theorem which states that,  in majoritarian group decisions, the errors in individual judgement are cancelled out to arbitrary accuracy as the number of voter increases.  This implies the stable dominance of rational majority in democratic decision making.  In reality, however, democratic majoritarianism can produce seemingly unexpected sudden shifts, as exemplified by political events such as the 2016 Brexit victory and Trump election.  There is also a glaring contradiction in democracy in which there exist powerful interest groups that exert disproportionate power despite their minority status.

Such unexpected and paradoxical outcomes of opinion dynamics are challenging puzzles, which are still to be elucidated. In particular with public opinion being nowadays  a major key to trigger eventual changes in modern societies as well as collapses of political regimes, the understanding of the mechanisms behind the making of public opinion becomes a major issue of vital interest.

It happened that from the beginning of sociophysics \cite{GGY82} the study of opinion dynamics has been a main topic of research, which has shed new and disruptive light about these paradoxical features resulting from majority group decision \cite{Bra,Ga12,Cast09,Sc18,Noor}. A great deal of models have been and are still proposed to describe opinion dynamics in social systems \cite{sznajd,och,sornette,frank-voter,neigbhor,sanchez}. Most of them consider homogeneous populations with a local update rule and discrete variables. A series of these models were shown to be included within a single unifying frame \cite{Ga05b}. Continuous variables have also been used \cite{DNAW00,HK02}. Yet, despite, ongoing active research \cite{RBA16,BCNBL16,MBG16,CM16,CG18,LPLB19}, Up-to-date the opinion dynamics issue is still lacking a comprehensive and robust framework especially a reliable predictive tool.

Among those models stands the seminal Galam model \cite{Ga86,GCMD98,Ga04a,Ga05a,FSE09}, which combines local majority rule updates with local symmetry breaking driven by unconscious prejudices and cognitive biaises in case of an even size group at a tie. The model has revealed some heuristic capacity with the successful predictions of a few political events such as 2016 Brexit victory \cite{Ga04a,Ga18},Trump election \cite{Ga17} and 2005 French rejection of the European constitution project \cite{LG05}. It has been subsequently extended to incorporate heterogeneous populations with three different psychological traits, which are respectively heterogeneous prejudices, inflexibility or stubbornness  \cite{GM91,GJ07,MG13,CO15,Mo15,PSL12,BRG16,JP18} and contrarianism  \cite{Ga04b,TM13,LHL17,GC17}. However, only specific combinations of these traits have been investigated \cite{ JG19,CBA14,KMT18}. An additional limitation has been the restriction of update groups to size 2, 3 and 4.

In this paper,  investigating further the Galam model in a generic parameter space of five dimensions we obtain a single universal update equation to follow the temporal evolution of opinion distribution among a heterogeneous population with any combination of rational, stubborn, contrarian agents for any average of hidden prejudices and for an arbitrary size of discussing groups.

The update equation allows exploring the complete geometry of the underlying landscape. In particular, attractors and tipping points, which shape the faith of possible dynamics flows, are unveiled. The underlying dynamics path of public opinion driven by repeated discussions among small groups of people during a social or political public campaign becomes predictable. Future occurrence of sudden minority spreading and parallel majority collapse ahead of their actual occurrence are revealed as shown with 2016 Brexit and Trump victories. They became visible within a forecast time horizon and in turn predictable. 

In addition, having the full five dimensional landscape makes predictions more feasible and robust since then the precise evaluation of the parameter values is not required. Instead, what matters is identifying the relevant basin of attraction where the dynamics will take place. However, along the basin boundaries more precise estimates of the parameters are necessary.

Having a universal formula which accounts for generic psychological features makes the model applicable to a large spectrum of social situations including voting outcomes, market shares and societal trends. The results allow also to envision novel winning strategies in competing environments to win a public debate or a market share.

The rest of paper is organized as follows. 
In section 2, we outline the derivation of the universal formula aggregating the various components of Galam model. A few illustration of sudden and unexpected minority outbreaks yielded by the equation are exhibited in Section 3, 
Section 4 contains some explicit expressions for the evolution equation for (some) FEW specific values of the group size $r$. 
We analyze the mathematical structure of the universal update equation through the identification of its fixed points in Section 5. Section 6 explores some aspects of the five-dimensional parameter space of the the model pointing out the existence of critical points.  The results are illustrated with several numerical examples. Concluding remarks are presented in the last Section.

\section{The universal formula in five dimensional parameter space}

We consider a population with $N$ agents each capable of taking two states $1$ and $0$, respectively  representing two exclusive opinions $A$ and $B$.  At a given time $t$ the corresponding proportions of agents holding A and  B are denoted $p_t$ and $(1-p_t)$. To make legitimate the use of proportions and probabilities we focus on cases with $N > 100$  \cite{FSE09}. Choosing an initial time $t=0$, we investigate the time evolution of $p_0$ driven by informal local discussions among agents at successive discrete time steps  with $p_{0} \rightarrow p_{1} \rightarrow p_{2} \rightarrow ...$ 

To account for this complicated and unknown process we use the Galam Dynamical Model, which monitors the opinion dynamics by a sequential iterated scheme. To implement one scheme, agents are first randomly distributed within small groups of size $r$. Then, within each group a majority rule is applied to update locally agent opinions. Each agent has one vote to determine the actual local majority. However, to account for the heterogeneity of psychological traits among agents, not all of them obey to the majority rule by shifting opinion if holding the local minority choice. Three types of agents, floaters, inflexibles and contrarians, are considered. After local updates are performed, all groups are dismantled and agents are reshuffled thus erasing the local correlations which were created by applying the local majority rules.
\begin{itemize}

\item
The {\em floaters} vote to determine the local majority and afterwards the ones holding the minority opinion shift to adopt the majority opinion.
However, in case of a tie at a local even group, floaters adopt the choice in tune with the leading prejudice among the group members  \cite{Ga12}. This prejudice driven tie breaking is activated unconsciously by the agents. Accordingly, at a tie opinion A is chosen with probability $k$ and opinion B with probability $(1-k)$. The value of $k$ satisfies $0\leq k \leq 1$ and is a function of the distribution of prejudices within the social group. 
\item
The {\em inflexibles} also vote to determine the local majority but contrary to the floaters,  in case they hold the minority choice they stick to it. They are stubborn and keep on their own pre-decided choice whatever the local majority is. Accordingly an $A$-inflexible stays in state $1$ and a $B$-inflexible in $0$.  The same holds true at a tie.  Their respective proportions denoted $a$ and $b$ are external fixed parameters with the constraints $0 \leq a+b \leq 1$.
\item
The {\em contrarians} are floaters, who once groups are dismantled, decide individually to shift opinion to oppose the group majority they contributed to. The attitude is independent of the majority choice, either $A$ or $B$.  The proportion of contrarians among floaters is denoted $c$ satisfying $0 \leq c \leq \ 1$. The value of $c$ is also a fixed external parameter  with the case $c > \frac{1}{2}$ corresponding to a situation where a majority of floaters systematically flips to the other opinion creating an on going alternative shifting between $A$ and $B$.  The proportion of floaters $A$ and $B$ being $(1-a-b)$, the proportion of contrarians among all agents is given by ${\tilde c}  = (1-a-b)c$.
\end{itemize}

At time $t$ once a scheme is completed with $p_{t} \rightarrow p_{t+1}$, another one is performed to get $p_{+1} \rightarrow p_{t+2}$ and so forth till the debate ends with an eventual vote at a given time, which determines the total number of updates to be implemented. The frequency of updates is a function of the intensity of the on going debate. For any real situation the debate initial proportion $p_0$ is evaluated using polls.

Let us now evaluate the function $P^{(r)}_{a, b, c, k}$ yielding $p_{t+1}$ from $p_{t}$ given the values of $(a, b, c, k)$ with,
\begin{eqnarray}
\label{e11}
p_{t+1} = P^{(r)}_{a, b, c, k} (p_t) .
\end{eqnarray}

However, since the contrarian phenomenon is mathematically identical to a random flipping of floaters with probability $c$ among reshuffled agents, we can decouple the $c$ effect from the local update writing, 
\begin{eqnarray}
\label{e112}
p_{t+1} = (1-c) \left[ P^{(r)}_{a, b, k} (p_t)-a\right] + c\left[1-P^{(r)}_{a, b, k} (p_t)-b\right].
\end{eqnarray}
which simplifies as,
\begin{eqnarray}
\label{e112-2}
p_{t+1} = (1-2c) P^{(r)}_{a, b, k} (p_t) + c(1+a-b).
\end{eqnarray}

To evaluate $P^{(r)}_{a, b, k}$ we note that in a given configuration of a group of size $r$ with $(r-\mu)$ agents holding opinion $A$ and $\mu$ agents holding opinion $B$,
the contributions to  $P^{(r)}_{a, b, k}$ result from two different families. One family includes contributions with a majority of $A$ and equality of $A$ and $B$ (for even size at a tie with probability $k$) while the other family corresponds to contributions with a minority of $A$ and equality of $A$ and $B$ (for even size at a tie with probability $(1-k)$). These families are denoted respectively $P^{(r,\mu)}_{a, b, c}$ and $Q^{(r,\mu)}_{a, b, c}$ under the constraint  $\mu \le [\frac{r}{2}]$ where $[x]$ is the integer part of $x$.
Eq. (\ref{e11}) can be rewritten as, 
\begin{eqnarray}
\label{e101}
P^{(r,\mu)}_{a, b, k} (p_t) = \sum_{\mu=0}^{[\frac{r}{2}]}  
\left\{ K^{(r,\mu)}_k P^{(r,\mu)}_{a, b}(p_t) + K^{(r,\mu)}_{1-k} Q^{(r,\mu)}_{a, b} (p_t)  \right\}
\end{eqnarray}
with,
\begin{eqnarray}
\label{e102}
K^{(r, \mu)}_k = 1+ (k-1) \delta_{r,2\mu} &\!\! = 1 \quad(r \ne2 \mu)
\nonumber \\ & = k  \quad(r = 2 \mu),
\end{eqnarray}
which allows to factorize the $k$ dependence.

\begin{figure}[h]
\begin{center}
\includegraphics[width=5cm]{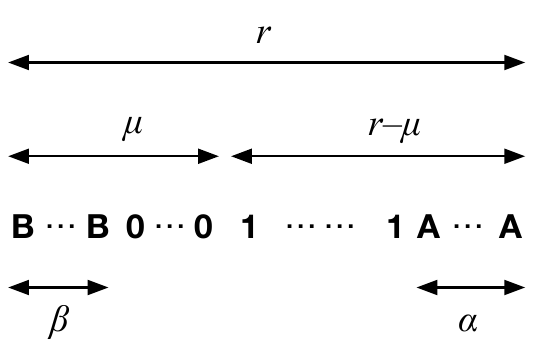}\qquad
\includegraphics[width=5cm]{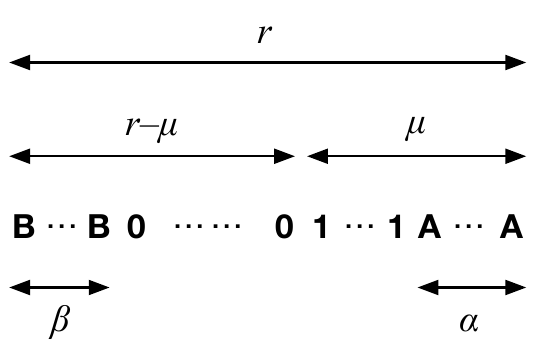}
\end{center}
\caption{
Dagrams representing a group of $r$ agents made up of $\alpha$ $A$-inflexibles (shown as $\bf{A}$), $\beta$ $B$-inflexibles (shown as $\bf{B}$), and $r^\alpha-\beta$ floaters and contrarians.  Agents are split into two opinions with $\mu$ having one opinion and $(r-\mu)$, the other with $\mu < \left[ \frac{r}{2} \right]$.  
The left diagram is the case in which  opinion $A$ holds the majority, and the right diagram is the case in which $B$ holds the majority.
 }
\label{f01}
\end{figure}

The quantity $P^{(r,\mu)}_{a, b}(p)$, which accounts for the contribution to $A$ after one update from configurations with $(r-\mu)$ agents with opinion  $A$  and $\mu$ agents with opinion $B$, can be decomposed as the sum of contributions with $\alpha$ $A$-inflexibles and $\beta$ $B$-inflexibles, where integer $\alpha$ runs from $0$ to $(r-\mu)$ and $\beta$, from $0$ to $\mu$ giving (See Fig.\ref{f01} left),
\begin{eqnarray}
\label{e114p}
P^{(r,\mu)}_{a, b}(p) = \sum_{\alpha=0}^{r-\mu} \sum_{\beta=0}^{\mu} 
P^{(r,\mu)}_{a, b, \alpha, \beta}(p)
\end{eqnarray}
Since $\mu \le [\frac{r}{2}]$ ($A$ majority or tie), contributions to $p_{t+1}$ from Eq. (\ref{e114p})  come  from all agents except $B$-inflexibles minus contrarians after the update yielding,
\begin{eqnarray}
\label{e103p}
P^{(r,\mu)}_{a, b, \alpha, \beta}(p)
= \begin{pmatrix} r \\ \mu \end{pmatrix} 
\begin{pmatrix} \mu \\ \beta\end{pmatrix} \begin{pmatrix} r-\mu \\ \alpha \end{pmatrix}
(1-p-b)^{\mu-\beta} 
\nonumber \\
\times (p-a)^{r-\mu-\alpha} 
b^\beta a^\alpha 
\left( \frac{r-\beta}{r} \right) .
\end{eqnarray}
Similarly, the quantity $Q^{(r,\mu)}_{a, b}(p)$, which accounts for the contribution to $A$ from configurations with $\mu$ opinion $A$
and $(r-\mu)$ $B$,
can be decomposed as the sum of contributions with  $\alpha=0,...,\mu$
$A$-inflexibles and  $\beta=0,...,(r-\mu)$
$B$-inflexibles giving (See Fig.\ref{f01} right),
\begin{eqnarray}
\label{e114q}
Q^{(r,\mu)}_{a, b}(p) = \sum_{\alpha=0}^{\mu} \sum_{\beta=0}^{r-\mu} 
Q^{(r,\mu)}_{a, b,\alpha, \beta}(p)
\end{eqnarray}
Since $\mu \le [\frac{r}{2}]$ ($A$ minority or tie), contributions to $p_{t+1}$ from Eq. (\ref{e114q})  come only from $A$-inflexibles 
after the update yielding,
\begin{eqnarray}
\label{e103q}
Q^{(r,\mu)}_{a, b, \alpha, \beta}(p)
= \begin{pmatrix} r \\ \mu \end{pmatrix} 
\begin{pmatrix} \mu \\ \alpha \end{pmatrix} \begin{pmatrix} r-\mu \\ \beta \end{pmatrix}
(1-p-b)^{r-\mu-\beta} 
\nonumber \\
\times
(p-a)^{\mu-\alpha} b^\beta a^\alpha 
 \left( \frac{\alpha}{r} \right) .
\end{eqnarray}
With the use of  the formulae
$(x+y)^n = \sum_{j=0}^n \begin{pmatrix} n \\ j \end{pmatrix} x^{n-j} y^j $
and 
$n y (x+y)^{n-1} = \sum_{j=0}^n \begin{pmatrix} n \\ j \end{pmatrix} j x^{n-j} y^j $, 

the summation over $\alpha$ and $\beta$ leads us to, 
\begin{eqnarray}
\label{e112-3}
P^{(r,\mu)}_{a, b}(p) =
\begin{pmatrix} r \\ \mu \end{pmatrix} p^{r-\mu}(1-p)^{\mu} 
\left[ 1 -\frac{\mu}{r (1-p)} b  \right] ,
\end{eqnarray}
and
\begin{eqnarray}
\label{e111}
Q^{(r,\mu)}_{a, b}(p) =
\begin{pmatrix} r \\ \mu \end{pmatrix} p^{\mu}(1-p)^{r-\mu} 
\left[ \frac{\mu}{r p} a \right] ,
\end{eqnarray}

With the use of identities
\begin{eqnarray}
\label{e115}
\sum_{\mu=0}^{[\frac{r}{2}]} 
K^{(r,\mu)}_{1-k}  
\begin{pmatrix} r \\ \mu \end{pmatrix} p^\mu(1-p)^{r-\mu}
\qquad\qquad\qquad\qquad
\nonumber \\
+
\sum_{\mu=0}^{[\frac{r}{2}]} 
K^{(r,\mu)}_{k}  
\begin{pmatrix} r \\ \mu \end{pmatrix} p^{r-\mu}(1-p)^{\mu}
\qquad
\nonumber \\
=(p + 1-p)^r=1 ,
\end{eqnarray}
and
\begin{eqnarray}
\label{e116}
\sum_{\mu=0}^{[\frac{r}{2}]} K^{(r,\mu)}_k  
\begin{pmatrix} r \\ \mu \end{pmatrix} \frac{\mu}{r p} p^\mu(1-p)^{r-\mu}
\qquad\qquad\qquad\qquad
\nonumber \\
+\sum_{\mu=0}^{[\frac{r}{2}]} K^{(r,\mu)}_{1-k} 
\begin{pmatrix} r \\ \mu \end{pmatrix} \frac{r-\mu}{r p} p^{r-\mu}(1-p)^{\mu}
\quad
\nonumber \\
=\frac{1}{rp}\left.\partial_\eta (\eta p+1-p)^r \right|_{\eta=1}
=(p + 1-p)^{r-1}=1 ,
\end{eqnarray}
we arrive at the compact expression of the evolution function $P^{(r)}_{a, b,k}$ for general group size $r$ in the form,
\begin{eqnarray}
\label{e200-o}
P^{(r)}_{a, b, k}(p)
&=&\!\!
 \Pi^{(r)}_0 (p,k) + a\, \Pi^{(r)}_1 (p,k) -b\, \Pi^{(r)}_2 (p,k) ,
\quad
\end{eqnarray}
where
\begin{eqnarray}
\label{e211}
\Pi^{(r)}_0 (p,k) =\sum_{\mu=0}^{[\frac{r}{2}]} \begin{pmatrix} r \\ \mu \end{pmatrix}
 K^{(r,\mu)}_k 
 p^{r-\mu}(1-p)^{\mu} 
\end{eqnarray}
\begin{eqnarray}
\label{e212}
\Pi^{(r)}_1 (p,k) =  \sum_{\mu=0}^{[\frac{r}{2}]} \begin{pmatrix} r \\ \mu \end{pmatrix} K^{(r,\mu)}_{1-k} 
 \frac{\mu}{r p} p^{\mu}(1-p)^{r-\mu} 
\end{eqnarray}
\begin{eqnarray}
\label{e213}
\Pi^{(r)}_2 (p,k) = \sum_{\mu=0}^{[\frac{r}{2}]} \begin{pmatrix} r \\ \mu \end{pmatrix} K^{(r,\mu)}_k 
\frac{\mu}{r (1-p)} p^{r-\mu}(1-p)^{\mu} .
\end{eqnarray}
Note the relation $\Pi^{(r,\mu)}_2 (p,k) =\Pi^{(r,\mu)}_1 (1-p, 1-k)$.  
Also note the relation
\begin{eqnarray}
\label{e113}
 P^{(r,\mu)}_{a, b}(p) + Q^{(r,\mu)}_{b,a}(1-p)=  \begin{pmatrix} r \\ \mu \end{pmatrix} p^{r-\mu} (1-p)^{\mu} .
\end{eqnarray}
which guarantees the covariance of  $P^{(r)}_{ a, b, k}(p)$ with respect to the renaming of opinions $A$ and $B$,
\begin{eqnarray}
\label{e113v}
P^{(r)}_{ a, b, k}(p)+ P^{(r)}_{b, a, 1-k}(1-p)=  1.
\end{eqnarray}

At this stage to get the complete update Equation we use Eq. (\ref{e112-2}) to get,
\begin{eqnarray}
\label{e200}
&&
\!\!\!\!\!\!\!\!\!\!\!\!\!\!\!\!\!\!\!\!
P^{(r)}_{a, b, c,k}(p)
\nonumber\\
&&
\!\!\!\!\!\!\!\!\!\!
=
(1-2c) 
\left[  \Pi^{(r)}_0 (p,k) + a\, \Pi^{(r)}_1 (p,k) -b\, \Pi^{(r)}_2 (p,k) \right]  
\nonumber \\
&  &\!\!\!\!\!\!
+c (1+a-b) ,
\end{eqnarray}
which allows calculating the opinion dynamics for any set of parameters $(a,b,c,k)$ given any group size $r$ for any initial support $p_0$ for opinion $A$. We thus have calculated a universal formula to predict opinion dynamics in a parameter space of five dimensions. 

We now go one step further to make our equation further realistic by considering a distribution of size with $r=1, 2, ..., L$ where $L$ is the largest update size. Usually, in most social situations $L$ is around $4, 5$ or $6$. The size weight $w_r$ of groups of size $r$ has to be evaluated from observation under the constraint $\sum_{r=1}^{r=L} w_r=1$. Including the size $r=1$ allows to consider agents who do not take part in local discussions during each update. Then, using Eq. (\ref{e200}) the associated global proportion of A supporters becomes, 
\begin{eqnarray}
\label{e201}
&&
\!\!\!\!\!\!\!\!\!\!
\!\!\!\!\!\!\!\!
G^{\{L\}}_{a, b, c,k}(p)
\nonumber\\
&&
\!\!\!\!\!\!\!\!\!\!
\!\!\!\!\!\!\!\!
=
\sum_{r=1}^{r=L} w_r P^{\{r\}}_{a, b, c,k}(p)
\nonumber\\
&&
\!\!\!\!\!\!\!\!\!\!
\!\!\!\!\!\!\!\!
=
(1-2c) \sum_{r=1}^{r=L} w_r
\left[  
\Pi^{(r)}_0 (p,k) + a\, \Pi^{(r)}_1 (p,k) -b\, \Pi^{(r)}_2 (p,k) 
\right] +c (1+a-b) .
\label{gen}
\end{eqnarray}

In principle, Eq.(\ref{gen}) allows to contribute to the issue of connecting the topology of a given network and the dynamics occurring through the network \cite{net}. Indeed, the set of values of weights $w_r$ and $L$ could then be extracted from the network.

\section{A few illustrations of sudden and unexpected minority outbreaks }
\begin{figure}[h]
\begin{center}
\includegraphics[width=5cm]{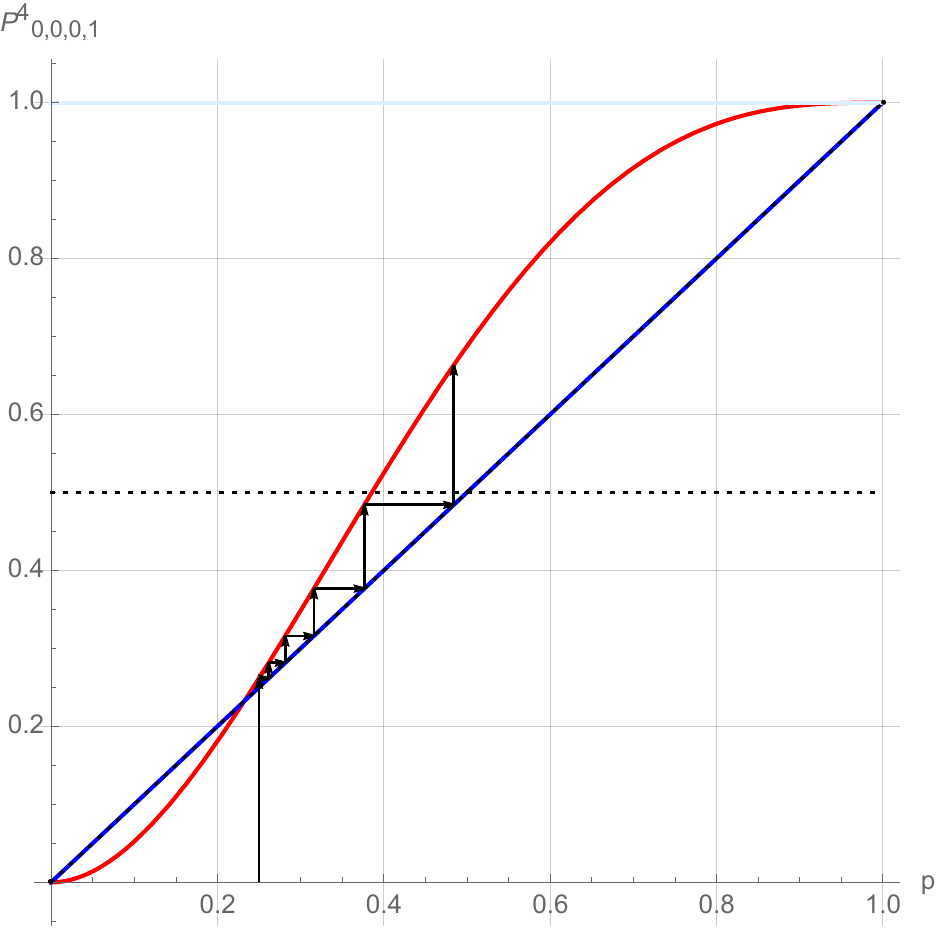}
\includegraphics[width=5cm]{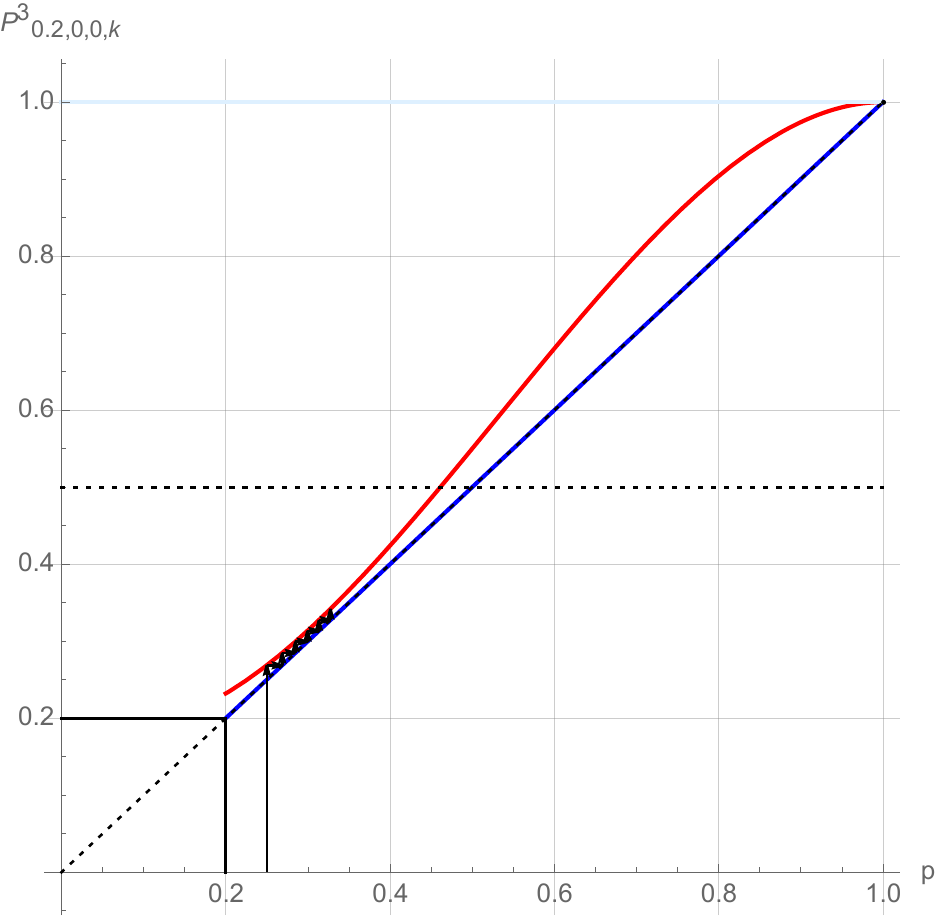}
\end{center}
\caption{
Left: Evolution of the proportion of A support $P^{\{4\}}_{0,0,0,1}(p)$ as a function of $p$. The arrows shows the minority spreading starting from an initial support $p=0.25$.
Right: Evolution of the proportion of A support  $P^{\{3\}}_{0.2,0,0,k}(p)$ as a function of $p$. The arrows shows the minority spreading starting from an initial support $p=0.25$.}
\label{c12}
\end{figure}

To make our universal update equation more concrete we list explicit expressions for the series of values $r=3,4,5,6,7,8,9$ in Appendix A. In addition, to distinguish the dynamical effects that come from the current update equation from the ones already present in the former model, we recall that the impact of unconscious prejudices has been studied for group size $r=4$ and the effect of having inflexible or stubborn agents for groups of size $3$. Each yields separately the phenomenon of minority spreading as exhibited for instance with $k=1$ and $a=0.20, b=0$ in Figure \ref{c12}.
Corresponding update formulas are recovered from Eq.(\ref{e200}) with $P^{(4)}_{0, 0, 0, 1}(p)=\Pi^{(4)}_0(p,k) =p^2(3p^2-8p+6)$ and $P^{(3)}_{0.20, 0, 0 ,k}(p)=\Pi^{(3)}_0(p,k) +0.20 \Pi^{(3)}_1(p,k) =-2p^3+3.20 p^2+0.20(1-2p)$. Notice that $P^{(3)}_{0.20, 0, 0 ,k}(p)$ is independent of $k$ as expected for an odd size group. Minority spreading occurs in both cases, but rather slowly for the second one.

\begin{figure}[h]
\begin{center}
\includegraphics[width=5cm]{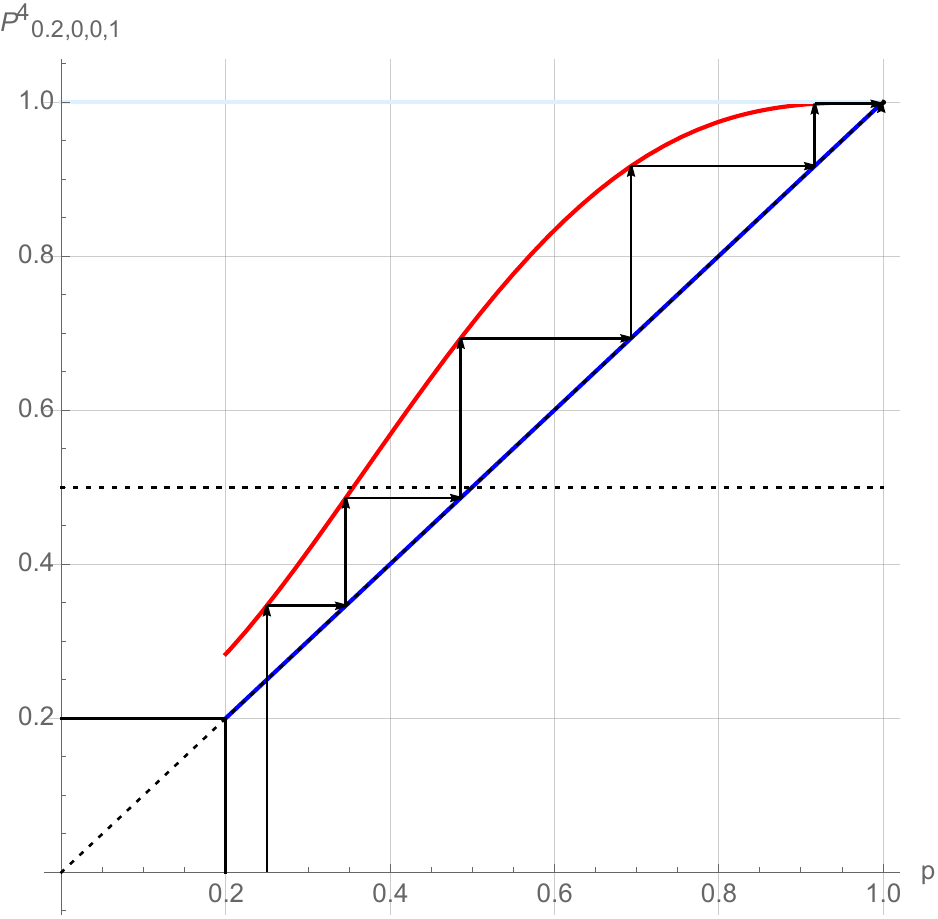}
\includegraphics[width=5cm]{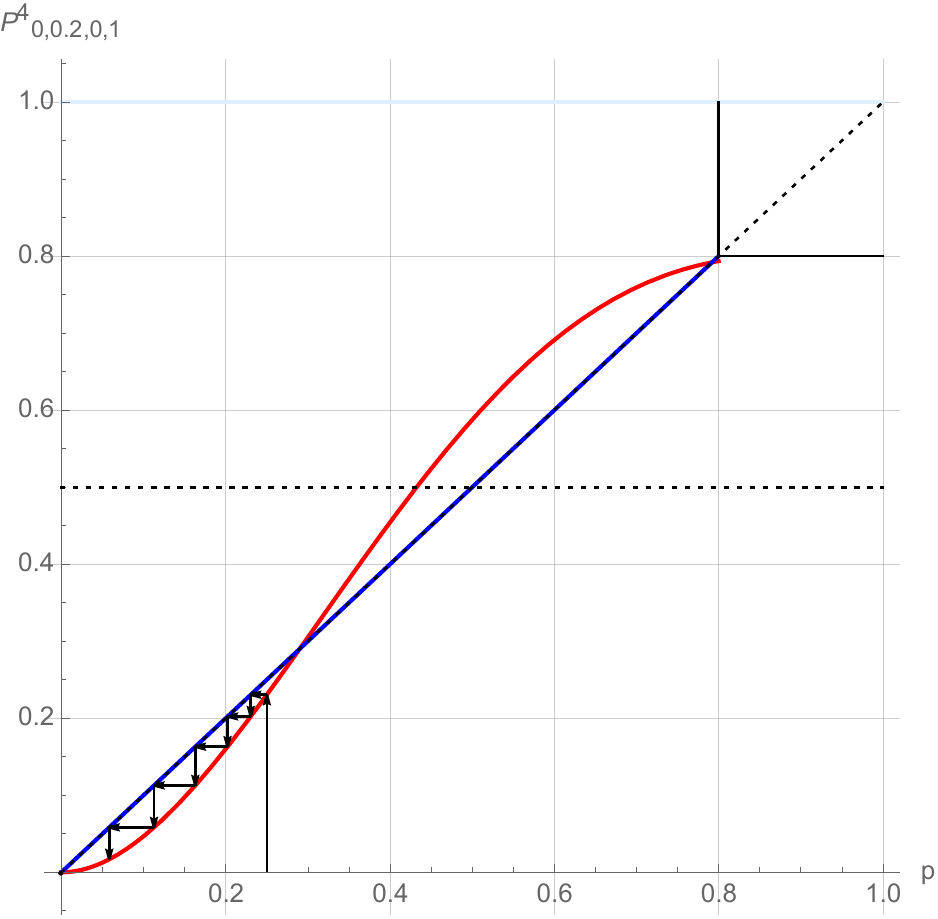}
\end{center}
\caption{
Left: Evolution of the proportion of A support $P^{\{4\}}_{0.20,0,0,1}(p)$ as a function of $p$. The arrows shows the speeding up of the A minority spreading starting from an initial support $p=0.25$.
Right: Evolution of the proportion of A support  $P^{\{4\}}_{0,0.2,0,k}(p)$ as a function of $p$. The arrows shows the A minority now shrinking from an initial support $p=0.25$.
}
\label{c34}
\end{figure}

While each effect was accounting for alone, now the universal formula allows to investigate the combination of both effects, either going along the same support for A ($P^{(4)}_{0.20, 0, 0, 1}(p)$) or competing for and against A ($P^{(4)}_{0, 0.20, 0, 1}(p)$) as shown Figure \ref{c34} for the case $r=4$. Combination enhances drastically the minority spreading but competition shows that $20\%$ of B stubborn do overcome the full benefit of the prejudice effect.

We proceed illustrating the capacity of the current update equation to yield sudden and unexpected minority outbreaks exhibiting a couple of striking cases with groups of size $r=3$ and $r=4$. 

\begin{figure}[h]
\begin{center}
\includegraphics[width=7cm]{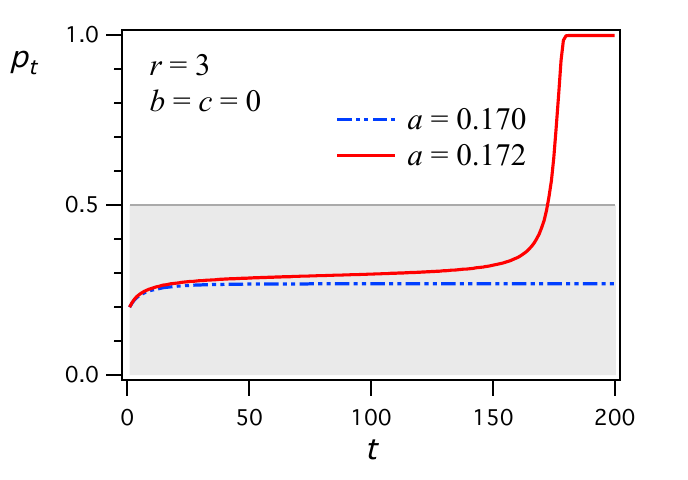}
\includegraphics[width=7cm]{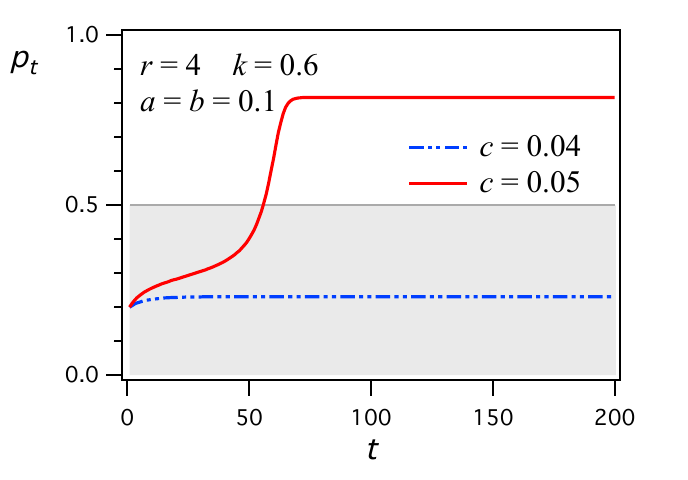}
\end{center}
\caption{
Evolution of $p$, the proportion of the support for A , as a function of the number of updates $t$. 
Left: A $r=3$ case with starting value $p=0.2$. Two lines are those of $a=0.170$ and $ a=0.172$, and common values $b= c=0$.
Right: A $r=4$ case with starting value $p=0.2$. Two lines are those of $c=0.04$ and $ c=0.05$, common values $a = b = 0.10$ and $k=0.6$.
}
\label{s1-7}
\end{figure}

In Figure \ref{s1-7}, each graph shows two curves for the evolution of $p$, the proportion of the support for A , as a function of the number of updates with the same value of $r$.  The two sets of parameter values are shown in each Figure.  Only very little differences differentiate the two sets of parameters while the two associated curves exhibit drastic difference in their respective outcomes.
The graph in the left is an example of $r=3$ case with the common starting value $p=0.2$. Two lines are those of $a=0.170$ and $ a=0.172$, and common values $b= c=0$, showing the drastic difference caused by a minuscule change in the parameter $a$.  Likewise, the graph in the right is an example of $r=4$ case with the common starting value $p=0.2$. Two lines are those of $c=0.04$ and $ c=0.05$, common values $a = b = 0.10$ and $k=0.6$, showing the effect caused by a $1\%$ change in the parameter $c$.

\section{Fixed points of the dynamics}

The dynamics implemented by repeated applications of Eq.(\ref{e200}) exhibits a rather large spectrum of different scenarios within the five dimensional space spanned by 
$0\leq a \leq p \leq 1$, $0\leq b \leq 1-p \leq 1$, $0\leq c \leq 1$, 
$a+b+c \leq1$, $0\leq k \leq 1$ and $1\leq r \leq \infty$. 
The phase diagram landscape is shaped by the various attractors and tipping surfaces, which are solutions of the fixed point equation,
\begin{eqnarray}
\label{e301}
p^\star = P^{(r)}_{a, b, c, k}(p^\star).
\end{eqnarray}

It is of interest to mention that Eq.(\ref{e301}) is a polynomial of degree $r$ in $p$ and thus exhibits $r$ solutions of which no more than three are real and contained within the $0-1$ range. This assessment results from playing with the equation and hand waving arguments but a mathematical proof is still on hold.

In case of three  fixed points $p_0^\star$, $p_s^\star$, $p_1^\star$, in ascending order. The smallest $p_0^\star \le \frac{1}{2}$ is stable and represents $B$-majoritarian final state, while the largest  $p_0^\star \ge \frac{1}{2}$, also stable, represents $A$ majority final state.  The medium valued $p_s^\star$ is unstable, and acts as a separator of the basins of attraction to $p_0^\star$ and $p_1^\star$, it is a tipping point of the dynamics. 
\begin{figure}[h]
\begin{center}
\includegraphics[width=5cm]{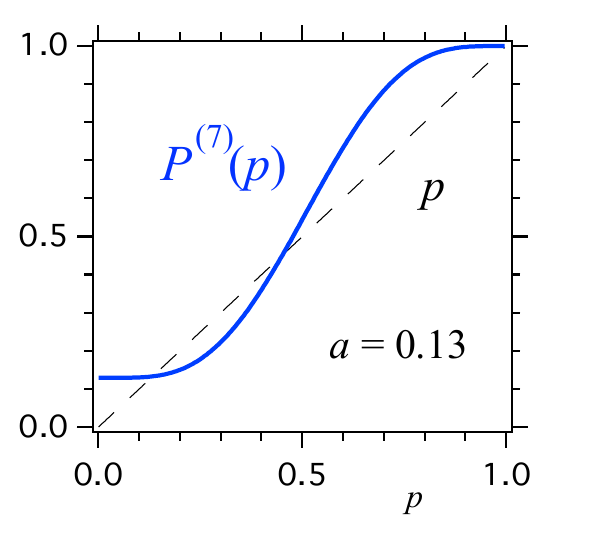}
\includegraphics[width=5cm]{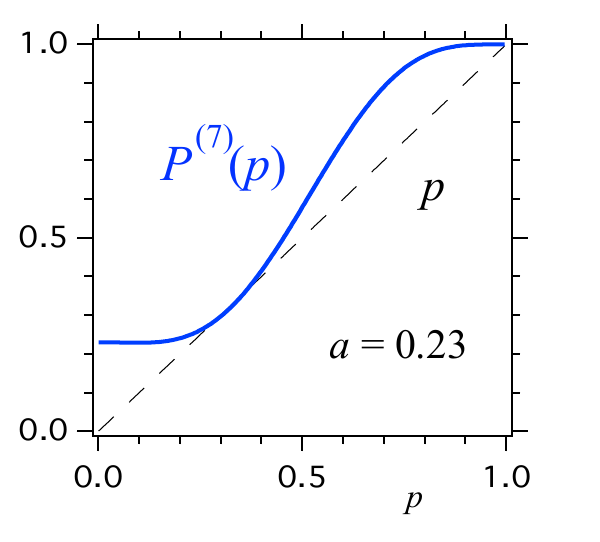}
\includegraphics[width=5cm]{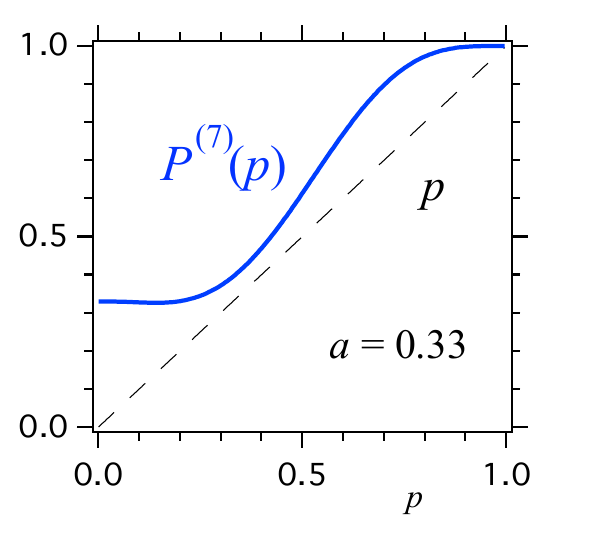}
\end{center}
\caption{
Examples of the evolution function $P^{r}_{a, b, c, k}(p)$ around critical value $a=a_c$ for $b=c=0$.  The group size is chosen to be $r=7$.
}
\label{f02}
\end{figure}
\begin{figure}[h]
\begin{center}
\includegraphics[width=5cm]{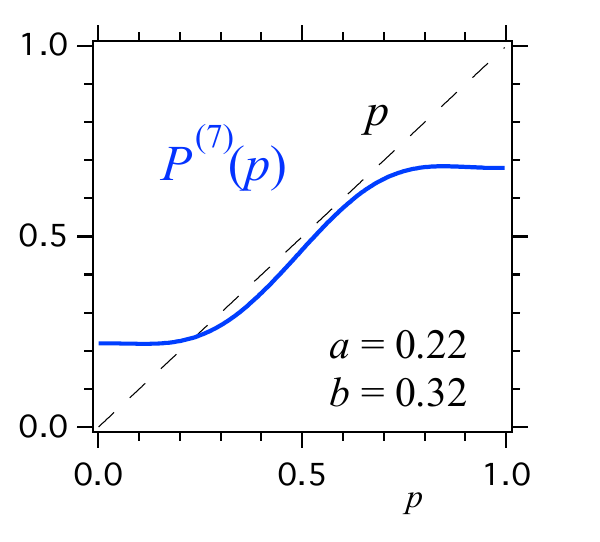}
\includegraphics[width=5cm]{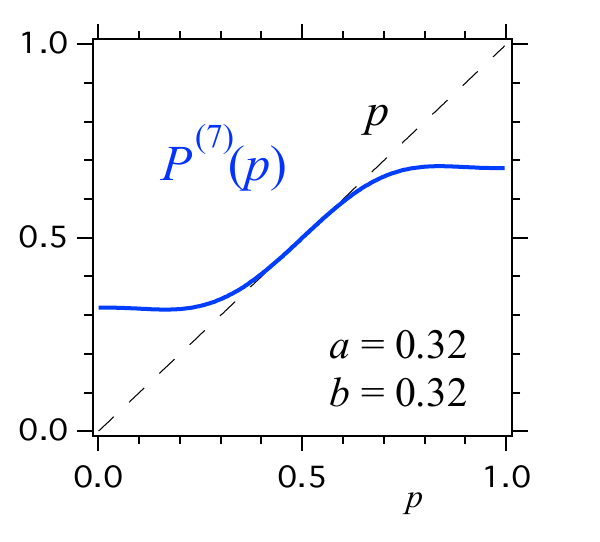}
\includegraphics[width=5cm]{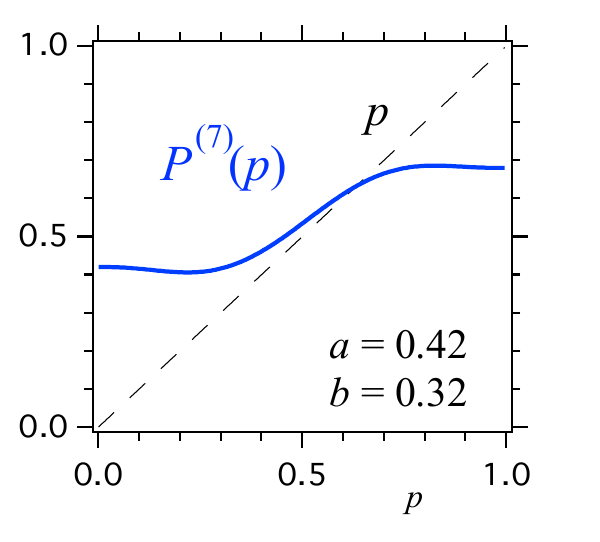}
\end{center}
\caption{
Examples of the evolution function $P^{r}_{a, b, c, k}(p)$ around triple critical value $a=b=a_t$ for $c=0$.  The group size is chosen to be $r=7$.
}
\label{f03}
\end{figure}

Although making a visual representation is impossible in five dimensions, being interested in the evolution of an initial value $p_0$ given fixed values of 
$(a,b,c,k, r)$, the operative use of the phase diagram is to select two-dimensional slices showing the evolution of $p_0$ as a function of repeated updates with $p_0 \rightarrow p_1 \rightarrow ... \rightarrow p_n$ where $n$ is the number of iterations. With respect to prediction about a real event, what matters is to determine if $p_n > \frac{1}{2}$ (opinion $A$ victory), $p_n < \frac{1}{2}$ (opinion $A$ failure) or $p_n \approx \frac{1}{2}$ (hung outcome).

An alternative practical use of the phase digram is to extract the two dimensional slices, which shows the function $p_{t+1}=P^{(r)}_{a, b, c, k}(p_t)$ for a fixed set of values $(a,b,c,k,r)$. These curves displays the eventual attractors and tipping points underlying the dynamics from which predictions can be made. These points are located at the crossing of $p_{t+1}=P^{(r)}_{a, b, c, k}(p_t)$ and the diagonal $p_{t+1}=p_{t}$.

Indeed all "slices" share the common property of having at least one single attractor for the dynamics. In addition, series of slices exhibits one additional attractor and a tipping point located between the two attractors. For those cases, varying some of the parameters $(a,b,c,k)$ may lead either to have one attractor and the tipping point to coalesce at critical values yielding then a single attractor dynamics. The other scheme is having the two attractors to merge at the tipping point to produce another single attractor dynamics. To have a single attractor dynamics implies one identified opinion is certain to win whatever its initial support is. That supposes the debate or the campaign duration lasts enough time to cross the $\frac{1}{2}$, i.e., $50\%$ of the ballots for an election.
Fig. \ref{f02} and Fig. \ref{f03} shows a series of illustrating cases of above two scenarios. 
%

\section{Exploring the phase diagram}

\begin{figure}[h]
\begin{center}
\includegraphics[width=7cm]{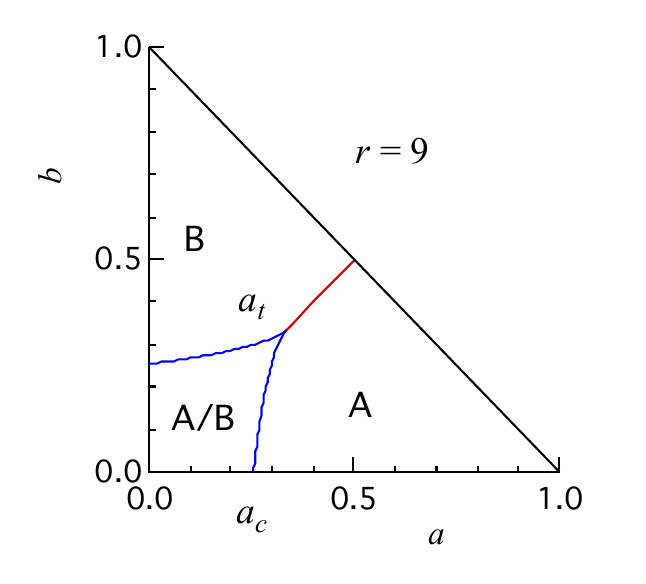}
\end{center}
\caption{
An example of the phase diagram of the parameter space $\{ a, b\}$.  The group size here is chosen to be $r=9$.   The regions marked as $A$, $B$, and $A/B$ represent parameter values with which the system converges unconditionally to $A$ majority, unconditionally to $B$ majority, and either $A$ or $B$ majority depending on the initial configuration, respectively.
}
\label{f04}
\end{figure}

From Eq.(\ref{e301}) when $b=c=0$, increasing the parameter $a$ makes $p_0^\star$ and $p_s^\star$ merge into single value at $a=a_c$, and then disappear to make $p_1^\star$ the sole final state of the system (See Fig.\ref{f04}). The critical value $a$ for $b=0$, which we call $a_c(c)$, is obtained from 
\begin{eqnarray}
\label{e311}
p^\star - P^{(r)}_{a_c(c), 0, c, k}(p^\star) = 0
\end{eqnarray}
\begin{eqnarray}
\label{e312}
1 - \partial_p P^{(r)}_{a_c(c), 0, c, k}(p^\star) = 0
\end{eqnarray}
In particular, for $c=0$, the critical value $a_c(0)$, which we simply call $a_c$, and its associated $p^\star_c$ are obtained froma
\begin{eqnarray}
\label{e313}
\Pi^{(r)}_0 (p^\star_c,k) \Pi^{(r)\prime}_1 (p^\star_c,k) -\Pi^{(r)\prime}_0 (p^\star_c,k) \Pi^{(r)}_1 (p^\star_c,k)
\qquad\qquad
\nonumber \\
+ \Pi^{(r)}_0 (p^\star_c,k)-p^\star_c\Pi^{(r)\prime}_0 (p^\star_c,k)= 0
\end{eqnarray}
\begin{eqnarray}
\label{e314}
a_c = \frac{p^\star_c-\Pi^{(r)}_0 (p^\star_c,k)}{\Pi^{(r)}_1 (p^\star_c,k)}
\end{eqnarray}

With small but non-zero $b$, the system goes through transition between two stable fixed point phase and single fixed point phase, when $a$ is varied, at a critical value higher than $a_c$.  
\begin{figure}[h]
\begin{center}
\includegraphics[width=7cm]{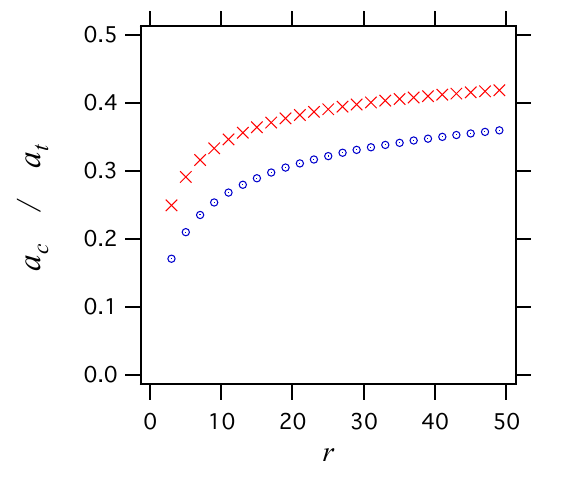}
\end{center}
\caption{
Critical parameters $a_c$ and $a_t$ as functions of group size $r$}
\label{f05}
\end{figure}
Above certain value of $b$, the system goes from $B$ majority single fixed-point phase to 
$A$ majority single fixed-point phase without passing through three fixed-point phase (See Fig.\ref{f04}).  The transition is characterized by triple critical point  $a=b=a_t(c)$
\begin{eqnarray}
\label{e321}
p^\star_t - P^{(r)}_{a_t(c), a_t(c), c, k}(p^\star_t) = 0
\end{eqnarray}
\begin{eqnarray}
\label{e322}
1 - \partial_p P^{(r)}_{a_t(c), a_t(c), c, k}(p^\star_t) = 0
\end{eqnarray}
limiting ourselves to $c=0$ for now, the triple point $a_t$$=a_t(0)$ and its associated $p^\star_t$ are obtained as $p^\star_t = \frac{1}{2}$ and
\begin{eqnarray}
\label{e324}
a_t = \frac{1-\Pi^{(r)\prime}_0 (\frac{1}{2},k)}{2 \Pi^{(r)\prime}_1 (\frac{1}{2},k)}
\end{eqnarray}

It is instructive to draw the phase diagram on $\{a, b\}$ plane.  
There is a region of small $a$ and $b$ in which final majority can go either way depending on their initial support.  For large value of $a$ and/or $b$, final majority is predetermined due to the strong influence of inflexibles.  When $a$ or $b$ exceeds $a_t$, two inflexible-dominated regions are placed next to each other without intermediate region of floater-determinability.  An example of $r=9$ is shown in Fig. \ref{f04}.

The $r$-dependence of $a_c$ and $a_t$ can be seen in Fig. \ref{f05}.  We list just some of them: 
For $r=3$, we have $a_c=0.1714$ and $a_t=0.25$, and for $r=5$,  $a_c=0.2104$ and $a_t=0.2917$.
For $r=7$, we have $a_c=0.2358$ and $a_t=0.3167$, and for $r=9$,  $a_c=0.2452$ and $a_t=0.3339$.

\begin{figure}[h]
\begin{center}
\includegraphics[width=7cm]{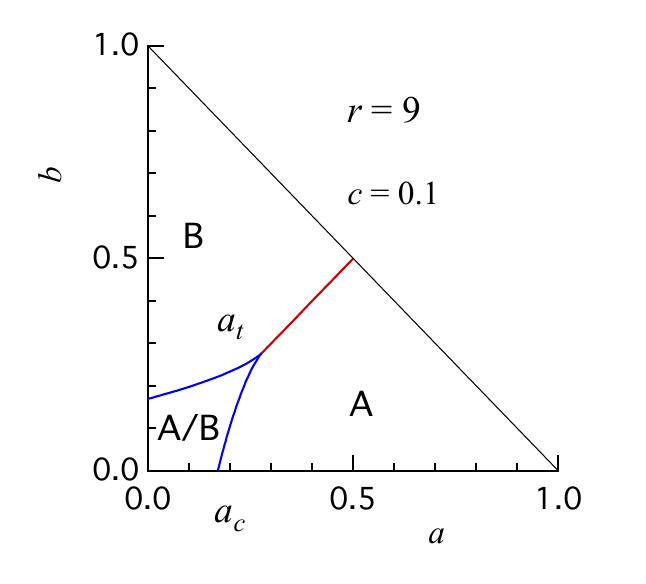}
\includegraphics[width=7cm]{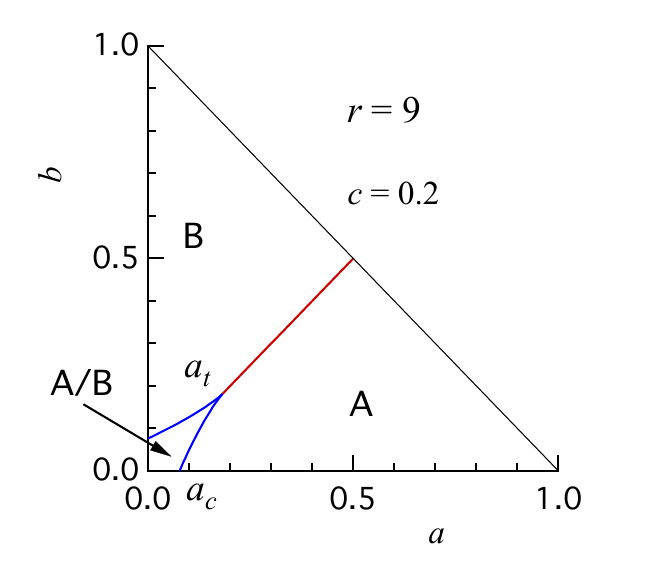}
\end{center}
\caption{
Phase diagram on $\{a, b\}$ plane for $r=9$.  Left for $c=0.1$ and right for $c=0.2$.  The regions marked as $A$, $B$, and $A/B$ represent parameter values with which the system converges unconditionally to $A$ majority, unconditionally to $B$ majority, and either $A$ or $B$ majority depending on the initial configuration, respectively.
}
\label{f06}
\end{figure}
Overall, approach to $r=\infty$ limit, $a_r=a_t=0.5$ is very slow, and for reasonably small group size, $r \approx 10$,  $a_c \approx \frac{1}{4}$ and $a_t \approx \frac{1}{3}$ holds, which are closer to $r=3$ case than $r=\infty$.   Even at $r\approx 50$, we have surprisingly small $a_c \approx \frac{1}{3}$ and $a_t \approx \frac{2}{5}$.

It is again instructive to look at the phase diagram on $\{a, b\}$ plane with different values of $c$ (Fig. \ref{f06}).  
It should be now very clear, that, for all group size $r$,  the effect of contrarians to the final majority formation is to decrease the role of floaters and increase the power of inflexibles. This fact, which has originally been found in $r=3$ example, turns out to be a generic feature of the Galam model.

\section{Summary}

In this paper, we have obtained a universal formula for the temporal evolution of agents following the Galam opinion dynamics in a parameter space of five dimensions, which are the respective proportions of each side inflexibles (stubbornness), the proportion of contrarians, the mean value of shared prejudices and the update group size.

The associated opinion landscape is found to be shaped by several attractors and tipping points, which yield a rich variety of non linearities and singularities in the opinion flows.  Sudden upheavals like minority spreading and majority collapse are thus given a rationale unveiling the hidden mechanisms behind the occurrence of unexpected and sudden shifts in the distribution of opinions like with the Brexit victory and Trump election. The time dependence of the corresponding phenomena is also exhibited.

For large group sizes, although the effects of inflexible agents waved out with simple majority rule holding, the results vary very slowly, showing earlier insights obtained from group sizes three and four remain intact. In contrast, for even sizes although the prejudice effect weakens with increasing size, it stays effective. Moreover, having the update equation for any group size $r$ makes possible to consider combinations of group sizes, which in turn opens the path to account for underlying networks.

The study of one specific case has showed the existence of critical points $a_c$ and $a_t$, which corner and branch out the lines separating the inflexibles-dominant and floaters-determinable phases.  The admixture and increase of contrarians effectively assist the dominance of inflexibles and reduce the parameter region in which initial composition of floaters can determine the outcome of majority opinion formation. The increase of group size $r$ induces modest but clear increase of both $a_c$ and $a_t$.  Accordingly, when the local discussions involve more agents, a larger committed minority is necessary to impose its will on the majority, or identically, a larger number of enlightened minority is necessary for its persuasion of the majority.

Getting the complete phase diagram of the dynamics of opinions yields a robust forecasting frame for which predictions become more reliable since what matters is the identification of the basin of attraction in which the dynamics is taking place thus providing some flexibility in the accuracy of parameters evaluation. However, along the basin boundaries, higher precision is required for the value of the parameters.  Given a specific issue, locating the corresponding relevant attractor allows anticipating a possible unexpected shift of opinion trend ahead of its occurrence.

It is of importance to emphasize that one instrumental feature of the model is that it does not rely on data beside the use of few polls. Indeed, the model is not data-driven, allowing to build the general landscape of the dynamics, whose topology is shaped by the various tipping points and attractors, which have been identified from the universal formula. 

However, at this stage, making the framework operational for robust predictions is still to be implemented since specific tools must be designed to evaluate the value of parameters, i.e., the proportions of each side inflexibles and contrarians, as well as the effective average of active prejudices connected to a given issues. This challenge cannot be addressed by physics solely, an interdisciplinary frame with social scientists is at stake to create effective new tolls to evaluate quantitatively those parameters. 

Yet, it has been shown that rough qualitative estimates can be sufficient to identify which basin is going to drive the dynamics for some cases as illustrated for the Brexit and French referendum about the project of European constitution for which most activated prejudices were acting in favor of the no. 

Accordingly, at this stage, predictions are qualitative about winning or losing a vote not about a precise value of the voting outcome. The forecast is about towards an increase of support to get above fifty percent or a decrease to end up below fifty percent. Therefore, interdisciplinary efforts have still to be done for reach solid quantitative predictions. Nevertheless, having the universal formula allows to elaborate winning strategies in competing environments widening previous discussed paths \cite{w1,w2,w3,w4,w5}.

To conclude, we have obtained a potentially ready-to-use universal formula which, although  operative tools to precise tuning of parameters are still lacking, provides already a new ground to make a large spectrum of predictions of winning strategies about outcomes of  opinion dynamics, including voting, market shares and societal trends. 

\section*{Acknowledgements}
The authors express deep gratitudes to Ms. Miwa Ohtsuki and Professor Shinichiro Inaba for their encouragements during the completion of this work. This manuscript has been released as a pre-print at arXiv  \cite{ar}.

\appendix
\section{Explicit evolution equations for $r=3,4,5,6,7,8,9$}
To make our universal update equation more concrete we list explicit expressions for the series of values $r=3,4,5,6,7,8,9$ with:\\
$r=3$
\begin{eqnarray}
\label{e253}
&&\!\!\!\!\!\!\!\!
\Pi^{(3)}_0(p,k) = p^2(3-2p)
\nonumber \\
&&\!\!\!\!\!\!\!\!
\Pi^{(3)}_1(p,k) = (1-p)^2
\nonumber \\
&&\!\!\!\!\!\!\!\!
\Pi^{(3)}_2(p,k) = p^2 ,
\end{eqnarray}
$r=4$
\begin{eqnarray}
\label{e254}
&&\!\!\!\!\!\!\!\!
\Pi^{(4)}_0(p,k) = p^3(4-3p) + 6 k p^2(1-p)^2
\nonumber \\
&&\!\!\!\!\!\!\!\!
\Pi^{(4)}_1(p,k) = (1-p)^2(1+2p) -3k p (1-p)^2
\nonumber \\
&&\!\!\!\!\!\!\!\!
\Pi^{(4)}_2(p,k) = p^3+3kp^2(1-p) ,
\label{r4}
\end{eqnarray}
$r=5$
\begin{eqnarray}
\label{e255}
&&\!\!\!\!\!\!\!\!
\Pi^{(5)}_0(p,k) = p^3(10-15p + 6 p^2)
\nonumber \\
&&\!\!\!\!\!\!\!\!
\Pi^{(5)}_1(p,k) = (1-p)^3( 1+3p)
\nonumber \\
&&\!\!\!\!\!\!\!\!
\Pi^{(5)}_2(p,k) = p^3 (4-3p) ,
\end{eqnarray}
$r=6$
\begin{eqnarray}
\label{e256}
&&\!\!\!\!\!\!\!\!
\Pi^{(6)}_0(p,k) = p^4(14-24p+10p^2) + 20 k p^3(1-p)^3
\nonumber \\
&&\!\!\!\!\!\!\!\!
\Pi^{(6)}_1(p,k) = (1-p)^3(1+3p+6p^2) - 10k p^2(1-p)^3
\nonumber \\
&&\!\!\!\!\!\!\!\!
\Pi^{(6)}_2(p,k) = p^4(5-4p) +10kp^3 (1-p)^2 ,
\end{eqnarray}
$r=7$
\begin{eqnarray}
\label{e257}
&&\!\!\!\!\!\!\!\!
\Pi^{(7)}_0(p,k) = p^4(35 -84p + 70 p^2-20p^3)
\nonumber \\
&&\!\!\!\!\!\!\!\!
\Pi^{(7)}_1(p,k) = (1-p)^4( 1+4p+10p^2)
\nonumber \\
&&\!\!\!\!\!\!\!\!
\Pi^{(7)}_2(p,k) = p^4 (15-24p+18p^2) ,
\end{eqnarray}
$r=8$
\begin{eqnarray}
\label{e258}
&&\!\!\!\!\!\!\!\!
\!\!\!\!\!\!
\Pi^{(8)}_0(p,k) = p^5(56-140p+120p^2-35p^3) + 70 k p^4(1-p)^4
\nonumber \\
&&\!\!\!\!\!\!\!\!
\!\!\!\!\!\!
\Pi^{(8)}_1(p,k) = (1-p)^4(1+4p+10p^2+20p^3) - 35k p^3(1-p)^4
\nonumber \\
&&\!\!\!\!\!\!\!\!
\!\!\!\!\!\!
\Pi^{(8)}_2(p,k) = p^5(21-35p+15p^2) + 35kp^4 (1-p)^3 ,
\end{eqnarray}
$r=9$
\begin{eqnarray}
\label{e259}
&&\!\!\!\!\!\!\!\!
\Pi^{(9)}_0(p,k) = p^5(126 -420p + 540 p^2 - 315p^3 +70p^4)
\nonumber \\
&&\!\!\!\!\!\!\!\!
\Pi^{(9)}_1(p,k) = (1-p)^5( 1+5p+15p^2+35p^3)
\nonumber \\
&&\!\!\!\!\!\!\!\!
\Pi^{(9)}_2(p,k) = p^5 (56-140p+120p^2-35p^3) ,
\end{eqnarray}


\begin{thebibliography}{99}


\bibitem{GGY82} S. Galam, Y. Gefen and Y. Shapir, 
Sociophysics: A new approach of sociological collective behaviour. description of a strike,
J. Math. Soc. {\bf 9}, 1--13 (1982).

\bibitem{Bra} R. Brazil, 
 The physics of public opinion, Physics World, January issue (2012).

\bibitem{Ga12} S. Galam, 
{\em Sociophysics: A physicist's modeling of psycho-political phenomena}, Springer, New York (2012).


\bibitem{Cast09} C. Castellano, S. Fortunato, and V. Loreto, 
Statistical physics of social dynamics, Rev. Mod. Phys. {\bf 81}, 591--646 (2009).

\bibitem{Sc18} F. Schweitzer, 
Sociophysics, Physics Today {\bf 71}, 
40--47 (2018).

\bibitem{Noor} H. Noorazar, Recent advances in opinion propagation dynamics: A 2020 Survey, The Eur. Phys. J. Plus {\bf 135}, Article number: 521 (2020).

\bibitem{sznajd} K. Sznajd-Weron and J. Sznajd, 
Opinion evolution in closed community,
Int. J. Mod. Phys. C {\bf 11}, 1157--1165 (2000).

\bibitem{och} R. Ochrombel, 
Simulation of Sznajd sociophysics model with convincing single opinions
Int. J. Mod. Phys. C {\bf 12}, 1091--1091 (2001).

 \bibitem{sornette} A. Corcos, J.-P. Eckmann, A. Malaspinas, Y. Malevergne and D. Sornette, 
 Imitation and contrarian behavior: hyperbolic bubbles, crashes and chaos,
 Quantitative Finance 2, 264--281 (2002).

\bibitem{frank-voter} L. Behera and F. Schweitzer,  
On spatial consensus formation: Is the Sznajd model different from a voter model?
Int. J. Mod. Phys. C {\bf 14}, 1331-1354 (2003).

\bibitem{neigbhor} C.J. Tessone, R. Toral, P. Amengual, H.S. Wio, and M. San Miguel,
Neighborhood models of minority opinion spreading,
Eur. Phys. J. B {\bf 39},  535--544 (2004).

\bibitem{sanchez} J.  R. Sanchez, A modified one-dimensional Sznajd model,
arXiv: cond-mat/0408518 (2004).

\bibitem{Ga05b} S Galam, 
Local dynamics vs. social mechanisms: A unifying frame,
 Euro. Phys. Lett. {\bf 70}, 
705--711 (2005).

\bibitem{DNAW00} G. Deffuant, D. Neau, F. Amblard and G. Weisbuch, Mixing beliefs among interacting agents, Advances in Complex Systems 3, 87-98 (2000)

\bibitem{HK02} R. Hegselmann R and U. Krause, Opinion dynamics and bounded confidence models, analysis, and simulation, Journal of Artificial Societies and Social Simulation 5(3), (2002).

\bibitem{RBA16} N. Rodriguez, J. Bollen, Y.-Y. Ahn, Collective Dynamics of Belief Evolution under Cognitive Coherence and Social Conformity, 
PLoS ONE {\bf 11}, 
e0165910(15pp) (2016).

\bibitem{BCNBL16} F. Battiston, A. Cairoli, V. Nicosia, A. Baule and V. Latora, Interplay between consensus and coherence in a model of interacting opinions, 
Physica D {\bf 323-324}, 
12-19  (2016).

\bibitem{MBG16}
G. A. Marsan, N. Bellomo and L. Gibelli, Stochastic evolutionary differential games toward a systems theory of behavioural social dynamics, Math.Models and Methods in App. Sciences Vol. {\bf 26}, 
1051-1093 (2016).

\bibitem{CM16} T. Cheon and J. Morimoto, 
Balancer effects in opinion dynamics, 
Phys. Lett. A {\bf 380}, 429--434 (2016).

\bibitem{CG18} T. Cheon and S. Galam, 
Dynamical Galam model, 
Phys. Lett. A {\bf 382}, 1509--1515 (2018).

\bibitem{LPLB19} C. W. Lynn, L. Papadopoulos,  D. D. Lee and D. S. Bassett, Surges of collective human activity emerge from simple pairwise correlations, 
arXiv:1803.00118v4 (2019).


\bibitem{Ga86} S Galam, 
Majority rule, hierarchical structures, and democratic totalitarianism: A statistical approach, 
J. Math. Psycho. {\bf 30}, 
426--434 (1986).

\bibitem{GCMD98} S Galam, B Chopard, A Masselot, M Droz, 
Competing species dynamics: Qualitative advantage versus geography, 
Eur. Phys. J. B {\bf 4}, 529--531 (1998).

\bibitem{Ga04a} S. Galam, 
The dynamics of minority opinion in democratic debate, 
Phys. A {\bf 336}, 56--62 (2004).

\bibitem{Ga05a} S. Galam, 
Heterogeneous beliefs, segregation, and extremism in the making of public opinions,  
Phys. Rev. E {\bf 71},  046123(5pp) (2005).

\bibitem{FSE09} G. Fasano, A  Sorato and A. Ellero, 
A modified Galam's model for word-of-mouth information exchange,  
Physica A {\bf 388}, 3901--3910 (2009).


\bibitem{Ga18} S. Galam, 
Are referendum a machinery to turn our prejudices into rational choices?  An unfortunate answer from sociophysics, in The Routlrdge Handbook to  Referendums and Direct Democracy, Eds. L. Morel and M. Qvortrup, Routledge Chap. 19,  334--347 (2018).

\bibitem{Ga17} S. Galam, 
The Trump phenomenon, an explanation from sociophysics, 
Int. J. Mod Phys. B {\bf 31}, 1742015(17pp)  (2017).

\bibitem{LG05} P. Lehir, 
Les math\'ematiques s'invitent dans le d\'ebat europ\'een,  (interview of S. Galam), Le Monde 26/02, 23 (2005).


\bibitem{GM91}  S. Galam and S. Moscovici, 
Towards a theory of collective phenomena: Consensus and attitude changes in groups, 
 Euro. J. Soc. Psychol. {\bf 21}, 49--74 (1991).
 
 \bibitem{GJ07} S. Galam and F. Jacobs, 
 The role of inflexible minorities in the breaking of democratic opinion dynamics,
Physica A {\bf 381}, 366--376  (2007).

\bibitem{MG13} A. Martins and S. Galam, 
Building up of individual inflexibility in opinion dynamics,
Phys. Rev. E {\bf 87}, 042807(8pp) (2013). 

\bibitem{CO15} N. Crokidakis,and P. M. C. de Oliveira, 
Inflexibility and independence: Phase transitions in the majority-rule model, 
Phys. Rev. E {\bf 92}, 062122(9pp) (2015).

\bibitem{Mo15}
M. Mobilia, 
Nonlinear q-voter model with inflexible zealots, 
Phys. Rev. E {\bf 92}, 012803(11pp) (2015).

\bibitem{PSL12}
W. Pickering, B. K. Szymanski and C. Lim, 
Analysis of the high dimensional naming game with committed minorities, 
Phys. Rev. E {\bf 93}, 052311(9pp)  (2016). 

\bibitem{BRG16} K. Burghardt, W. Rand and M. Girvan, 
Competing opinions and stubbornness: connecting models to data,  
 Phys. Rev. E {\bf 93}, 032305(14pp) (2016).

\bibitem{JP18} J. S. Juul  and M. A. Porter, 
Hipsters on networks: How a minority group of Individuals can lead to an anti-establishment majority, 
arXiv:1707.07187v2 (2018).

\bibitem{Ga04b} S. Galam, 
Contrarian deterministic effects on opinion dynamics: the hung elections scenario, Physica A {\bf 333}, 453--460  (2004).

\bibitem{TM13} S. Tanabe and N. Masuda, 
Complex dynamics of a nonlinear voter model with contrarian agents, 
Chaos {\bf 23}, 043136(6pp) (2013).

\bibitem{LHL17} E. Lee, P. Holme and S. H. Lee,  Modeling the dynamics of dissent, Physica A {\bf 486}, 262--272 (2017).

\bibitem{GC17} J. P. Gambaro and N.Crokidakis, The influence of contrarians in the dynamics of opinion formation, 
Physica A {\bf 486}, 465--472  (2017). 

\bibitem{JG19} F. Jacobs and S. Galam,
 Two-opinions-dynamics generated by inflexibles and non-contrarian and contrarian floaters,  
 arXiv:0803.3150v1 (2008).

\bibitem{CBA14} N. Crokidakis, V. H. Blanco and C. Anteneodo,  
Impact of contrarians and intransigents in a kinetic model of opinion dynamics, 
Phys. Rev. E {\bf 89}, 013310(8pp)  (2014). 

\bibitem{KMT18} N. Khalil, M. S. Miguel and R. Toral, 
Zealots in the mean-field noisy voter model,  
arXiv:1707.04087v2  (2018).

\bibitem{net} F. Battiston, G. Cencetti, I. Iacopini V. Latora, M. Lucas, A. Patania, J.-G.Young and G. Petri,  Networks beyond pairwise interactions: Structure and dynamics,  Physics Reports 874, 1-92 (2020).


\bibitem{w1} Wang Z, Xia C, Chen Z, Chen G. Epidemic Propagation With Positive and Negative Preventive Information in Multiplex Networks [published online ahead of print, 2020 Jan 13]. IEEE Trans Cybern. 2020, 2960605 (2020).

\bibitem{w2} Wang Z., Guo Q., Sun S. and Xia C., The impact of awareness diffusion on SIR-like epidemics in multiplex networks,  Applied Mathematics and Computation, V 349, 134-147 (2019).

\bibitem{w3} Li J.,  Wang J., Sun S. and Xia C., Cascading crashes induced by the individual heterogeneity in complex networks,  Applied Mathematics and Computation, V 323, 182-192 (2018).

\bibitem{w4} Javarone M.A., Network Strategies in the Election Campaigns , JSTAT, V2014 Ð P08013 (2014).

\bibitem{w5} Javarone M.A. and Squartini T., Conformism-driven phases of opinion formation on heterogeneous networks: The q-voter model case. JSTAT, V2015, P10002 (2015).

%
\bibitem{ar} S. Galam and T. Cheon, Tipping Point Dynamics: A Universal Formula, arXiv:1901.09622v1 (2019).



\end{thebibliography}

\end{document}